# Physical explanation of the universal "(separation)$^{-3}$" law found numerically for the electrostatic interaction between two protruding nanostructures

Richard G Forbes

Advanced Technology Institute & Department of Electrical and Electronic Engineering, University of Surrey, Guildford, Surrey GU2 7XH, UK

E-mail: r.forbes@trinity.cantab.net

**Abstract**

Two conducting nanostructures situated on a conducting base-plate, and subject to a common externally applied macroscopic electrostatic field, interact because their electrons are part of a common electron-thermodynamic system. Except at very small separations, the interaction reduces the apex field enhancement factor (FEF) of each nanostructure, by means of an effect that has been called "charge blunting". A parameter of interest is the fractional reduction ($-\delta$) of the apex FEF, as compared with the apex FEF for the same emitter when standing alone on the base-plate. For systems composed of two or a few identical post-like emitters, or regular arrays of such emitters, interaction details have been investigated both by methods based on numerical solution of Laplace's equation, and by the use of line-charge models. For post separations $c$ comparable with the post height $h$, several authors have shown that the variation of ($-\delta$) with $c$ is well described by formulae having exponential or quasi-exponential form. By contrast, explorations of the two-emitter situation using the "floating sphere at emitter-plane potential" (FSEPP) model have predicted that, for sufficiently large $c$-values, ($-\delta$) falls off as $c^{-3}$. Numerical Laplace-type simulations carried out by de Assis and Dall'Agnol (arXiv1711.00601v2) have confirmed this limiting $c^{-3}$ dependence for six different situations involving pairs of protruding nanostructures, and have led them to suggest that it may be an universal law. By using the FSEPP model for the central structure, and by adopting a "first moments" representation for the distant structure, this letter shows that a clear physical reason can be given for this numerically discovered general $c^{-3}$ dependence, for large $c$. An implication is that the quasi-exponential formula found applicable for $c\sim h$ is simply a good fitting formula, particularly in this range. A second implication is that the FSEPP model, which currently is used mainly in nanoscience, may have much wider applicability to electrostatic phenomena.

There continues to be interest in the use of large-area field electron emitters (LAFEs) as large-area electron sources. A particular recent driving force has been US military interest [1] in the development of advanced cold cathode electron emitter concepts for compact, high power, high frequency, vacuum electronic devices that would operate at a temperature of less than 1000 °C and provide a total emission current of at least 10 mA at an average (over area) current density of a few times $10^5$ A/m², for several hundred hours of continuous-wave operation. In this context and others, there has been interest (e.g. [2]) in LAFEs built using arrays of carbon-based post-like emitters, fabricated on a base plate.

Post-like emitters of this kind are particularly effective because they have a relatively high ratio of height to apex-radius, which causes the apex field enhancement factor (FEF) to be relatively high. The apex FEF $\gamma_a$ can be defined as the ratio ($E_a/E_M$) of the local electrostatic field $E_a$ at the emitter apex to the "macroscopic" electrostatic field $E_M$ would be present in the absence of emitting structures. In a commonly considered geometrical situation, where an isolated emitter stands on one of a pair of well-separated parallel plane plates of large lateral extent, $E_M$ is the mean electrostatic field between the plates, well away (laterally) from the emitter.

A relatively high FEF value means that the local field $E^{on}$ for onset of field electron emission (FE)—typically a few V/nm in magnitude—can be reached at a relatively low value of $|E_M|$ and hence at a relatively low value of the voltage applied between a counter electrode and the base plate on which the emitter stands. Use of relatively low applied voltages is technologically advantageous.

It is also well known that post-like emitters of this kind interact electrostatically and that (except at very small emitter separations) this interaction—sometimes called "shielding"—causes a reduction in apex-FEF values. This shielding effect is of technological interest, because it influences the inter-emitter spacing at which the macroscopic emission current density $J_M$ (the mean current density from a regular array of emitters) has its maximum value for a given system geometry and applied voltage.

Thus, there has been considerable interest in the basic electrostatics, both of large arrays of identical post-like emitters, and of pairs of identical emitters. Both numerical and analytical approaches have been used. A parameter of interest is the fractional reduction $(-\delta)$ in apex FEF, defined by

$$(-\delta) = -(\gamma_a - \gamma_{one})/\gamma_{one} \qquad (1)$$

where $\gamma_{one}$ is the apex FEF for a single isolated emitter, and $\gamma_a$ is the apex FEF in the two or many emitter situation. For regular arrays of identical emitters, with nearest-neighbour spacing $c$, the value of $(-\delta)$ for a given emitter will depend on the emitter shape, on $c$, and on the number and configuration of other emitters present. The simplest cases to analyse are a pair of identical emitters, and an infinite regular array of such emitters.

Early work on this problem used the so-called "HCP model", which models the emitter as a conducting hemisphere of radius $r$ on top of a conducting cylindrical post of radius $r$ and length $\ell$, the whole structure having height $h=\ell+r$. Analyses of arrays and clusters of HCP-model emitters (e.g., [3,4]), based on numerical solution of Laplace's equation, suggested that for post spacings $c$ comparable with the post height $h$, the variation of $(-\delta)$ with $c$ could be adequately represented by an exponential formula of the form

$$(-\delta) \sim \exp[a(c/h)] \qquad (2)$$

where $a$ is a constant with a value that depends on the details of the situation.

Subsequently, line-charge models (LCMs) for post-like emitters were introduced by Harris, Jensen and colleagues (e.g., [5]), and were used to examine electrostatic interactions between pairs of emitters and within small clusters of emitters. They found that slightly better fits could be generated by using an equation of the form

$$(-\delta) \sim \exp[a(c/h)^\kappa] \qquad (3)$$

where $\kappa$ is a constant found by a fitting procedure.

The "floating sphere at emitter plane potential" (FSEPP) model (e.g., [6]) is a simplified version of the HCP model, and allows simple analytical treatments of the interaction between two emitters, each represented as a floating sphere. With this model, it was found [6] that at sufficiently large distances the fractional reduction $(-\delta)$ falls off, not exponentially, but as the inverse-third power of distance, via the formula

$$(-\delta) \sim 2(r/\ell)(\ell/c)^3 . \qquad (4)$$

Since the FSEPP model is an approximate model, it seemed worthwhile to use numerical Laplace-type simulations to investigate the discrepancy between this $c^{-3}$ result and the earlier exponential or quasi-exponential fall-off. This has been done de Assis and Dall'Agnol [7], who have investigated six different structures (including the HCP model and the FSEPP model). In all cases they found $(-\delta)$ falls off as $c^{-3}$, and have suggested that this is an universal law. The purpose of this letter is to develop a basic physical argument for thinking that $c^{-3}$ dependence should be universal.

Reference [6] has shown that two physical effects contribute to apex-FEF changes: so-called "charge-blunting" and a neighbour-field effect, with charge-blunting usually the dominant effect, certainly at large separations. Physically, charge-blunting arises because the Fermi level must be constant throughout the system comprising the emitters and the base-plate (and electrically connected items at the same voltage), and must remain unchanged as the separation of two protruding nanostructures on the base-plate is varied. As the second nanostructure is moved closer to a "central" emitter, the charges on the second nanostructure generate, at the apex of the central emitter, an electrostatic potential $\delta\Psi$ of increasingly large magnitude. The electron system responds by moving charge between the central emitter and the system bulk, thereby reducing the charge magnitude near the emitter apex, and hence reducing the magnitudes of the apex field and apex FEF.

Using the FSEPP model for the central emitter, it is first shown that $(-\delta)$ is proportional to $\delta\Psi$, and hence to the conventional electrostatic field $E_0$ generated by the distant nanostructure (and its electrical image) at the point "0" where the axis of the central emitter intersects the base plane. For simplicity, only leading terms are used, and (as usual) it is assumed that all surfaces have the same local work-function.

In the absence of any distant nanostructure, the requirement that the electrostatic potential at the apex of the central emitter be equal to that immediately outside the base plate results in the formula

$$-E_M h + q_{one}/4\pi\varepsilon_0 r \approx 0 \tag{5}$$

where $E_M$ is the conventional macroscopic electrostatic field that would be present in the absence of the central emitter and $q_{one}$ is the charge formally placed at the centre of the sphere. Both $E_M$ and $q_{one}$ are negative for a field electron emitter. It follows that (when small terms are neglected) the conventional electrostatic field $E_{one}$ at the emitter apex is given by

$$E_{one} \approx q_{one}/4\pi\varepsilon_0 r^2 \approx E_M(h/r) , \tag{6}$$

and that the apex FEF $\gamma_{one}$ for the (isolated) central emitter is given adequately by:

$$\gamma_{one} = E_{one}/E_M \approx h/r . \tag{7}$$

Now consider that a distant nanostructure is present, and that this generates an additional electrostatic potential difference $\delta\Psi$ at the central emitter apex. As a result, the charge, apex field and apex FEF are changed by amounts $\delta q$, $\delta E_a$ and $\delta\gamma_a$, respectively, and equations (5) to (7) above are replaced by:

$$-E_M h + \delta\Psi + (q_{one} + \delta q)/4\pi\varepsilon_0 r \approx 0 , \tag{8}$$

$$E_a = E_{one} + \delta E_a \approx (q_{one} + \delta q)/4\pi\varepsilon_0 r^2 \approx E_M(h/r) - \delta\Psi/r , \tag{9}$$

where $E_a$ is the modified apex field, and

$$\gamma_a = E_a/E_M \approx (h/r) - \delta\Psi/E_M r . \tag{10}$$

It follows that the fractional reduction $(-\delta)$ in apex FEF is given adequately by:

$$(-\delta) = -(\gamma_a - \gamma_{one})/\gamma_{one} \approx \delta\Psi/E_M h . \tag{11}$$

In the lowest order of approximation, which will be adequate if the distant nanostructure is sufficiently far away (i.e., $c$ is sufficiently large), $\delta\Psi$ is given by

$$\delta\Psi \approx -E_0 h , \tag{12}$$

where $E_0$ is the conventional electrostatic field defined above (and is positive in value for FE). Hence

$$(-\delta) \approx -E_0/E_M . \tag{13}$$

Again, in the lowest order of approximation (which will be adequately valid for sufficiently large separations $c$), we can represent the charge distribution associated with the distant nanostructure (and its electrical image in the base plate) by the first moment of the charge distribution. This first moment will be a dipole of value $p$, situated in the base plane, a distance $c$ from the central emitter. If we define an "effective polarisability" $\alpha$ for the distant nanostructure, by $p = \alpha E_M$, then the field $E_0$ is given by

$$E_0 = -p/4\pi\varepsilon_0 c^3 = -\alpha E_M/4\pi\varepsilon_0 c^3 \tag{14}$$

And the final formula for $(-\delta)$ becomes

$$(-\delta) \approx \alpha/4\pi\varepsilon_0 c^3 . \tag{15}$$

It needs to be emphasised that this treatment is not, and is not intended to be, an exact analysis of what in practice can be very complicated geometrical situations. Rather, it is a "leading term analysis", the purpose of which is to bring out the underlying physics of the situation—in this case the underlying physics of the electrostatic interaction between a central emitter and a distant nanostructure, both situated on a conducting planar base-plate.

What formula (15) shows is that, whatever the geometrical form of the distant nanostructure, the fractional field reduction ($-\delta$) is expected to fall off with separation ($c$) as $c^{-3}$. This is the same result as found in the numerical calculations of de Assis and Dall'Agnol. The argument here supports/confirms their assertion that the $c^{-3}$ dependence is a universal result, by showing that there is an underlying physical reason for this form of dependence.

More generally, it should be noted that this is a result that is valid in the limit of large $c$. This result does not in itself conflict with the empirical finding that, in the range where $c \sim h$, numerical and LCM-model based results can be adequately fitted by an equation of form (3) above. A background issue has perhaps been whether the good performance of eq. (3) for $c \sim h$ was indicative of some good underlying physical basis that would allow eq. (3) to be physically valid for large $c$-values. The present work appears to show that this is *not* the case , and that eq. (3) is simply a good empirical fitting formula. Since it is a 3-parameter formula, it might reasonably be expected to have a good range of applicability.

A further point can perhaps usefully be made here. Although the FSEPP model [and the resulting approximate FEF-formula $\gamma_a \approx h/r$] is clearly an approximate model, it seems to be useful for generating results that (although approximate) display qualitatively correct physics. The numerical results of de Assis and Dall'Agnol [7] tend to support/confirm this. This suggests the possibility that the FSEPP model and associated formula $\gamma_a \approx h/r$ could perhaps have uses and valid applicability outside the realm of nanoscale physics in which (up till now) they have mainly been used.

**Acknowledgement**

I thank Drs de Assis and Dall'Agnoll for sending me a preliminary version [7] of their paper.

**ORCID iD**

Richard G Forbes "ID" https://orcid.org/0000-0002-8621-3298